\documentclass[twocolumn,pra,showpacs]{revtex4}
\usepackage{amsfonts}
\usepackage{amsmath}
\usepackage{graphicx}

\input{tcilatex}

\begin{document}

\title{Lindblad Rate Equations}
\author{Adri\'{a}n A. Budini$^{1,2}$}
\affiliation{$^{1}$Instituto de Biocomputaci\'{o}n y F\'{\i}sica de Sistemas Complejos,
Universidad de Zaragoza, Corona de Arag\'{o}n 42, (50009) Zaragoza, Spain\\
$^{2}$Consejo Nacional de Investigaciones Cient\'{\i}ficas y T\'{e}cnicas,
Centro At\'{o}mico Bariloche, Av. E. Bustillo Km 9.5, (8400) Bariloche,
Argentina}
\altaffiliation{present address}
\date{\today}

\begin{abstract}
In this paper we derive an extra class of non-Markovian master equations
where the system state is written as a sum of auxiliary matrixes whose
evolution involve Lindblad contributions with local coupling between all of
them, resembling the structure of a classical rate equation. The system
dynamics may develops strong non-local effects such as the dependence of the
stationary properties with the system initialization. These equations are
derived from alternative microscopic interactions, such as complex
environments described in a generalized Born-Markov approximation and
tripartite system-environment interactions, where extra unobserved degrees
of freedom mediates the entanglement between the system and a Markovian
reservoir. Conditions that guarantees the completely positive condition of
the solution map are found. Quantum stochastic processes that recover the
system dynamics in average are formulated. We exemplify our results by
analyzing the dynamical action of non-trivial structured dephasing and
depolarizing reservoirs over a single qubit.
\end{abstract}

\maketitle

\section{Introduction}

The description of open quantum systems in terms of\ local in time
evolutions is based in a weak coupling and Markovian approximations \cite%
{petruccione,cohen}. When these approximations are valid, the dynamics can
be written as a Lindblad equation \cite{petruccione,cohen,alicki,nielsen}.
The evolution of the density matrix $\rho _{S}(t)$\ of the system of
interest reads%
\begin{equation}
\frac{d\rho _{S}(t)}{dt}=\frac{-i}{\hbar }[H_{eff},\rho _{S}(t)]-\{D,\rho
_{S}(t)\}_{+}+F[\rho _{S}(t)],  \label{Lindblad}
\end{equation}%
where $H_{eff}$ is an effective Hamiltonian, $\{\cdots \}_{+}$\ denotes an
anticonmutation operation, and%
\begin{equation}
D=\frac{1}{2}\sum_{\alpha ,\gamma }a_{\alpha \gamma }V_{\gamma }^{\dagger
}V_{\alpha },\;\;\;\;\;\;\;F[\bullet ]=\sum_{\alpha ,\gamma }a_{\alpha
\gamma }V_{\alpha }\bullet V_{\gamma }^{\dagger }.
\end{equation}%
Here, the sum indexes run from one to $(\dim \mathcal{H}_{S})^{2},$ where $%
\dim \mathcal{H}_{S}$ is the system Hilbert space dimension. The set $%
\{V_{\alpha }\}$ corresponds to a system operator base, and $a_{\alpha
\gamma }$ denotes a semipositive Hermitian matrix that characterize the
dissipative time scales of the system.

Outside the weak coupling and Markovian approximations, it is not possible
to establish a general formalism for dealing with non-Markovian
system-environment interactions \cite%
{weiss,esposito,haake,garraway,tannor,ulrich,imamoglu,breuer}. Nevertheless,
there exist an increasing interest in describing open quantum system
dynamics in terms of non-Markovian Lindblad equations \cite%
{barnett,wilkie,wilkieChem,budini,cresser,lidar,sabrina,maniscalco,jpa,gbma,salo}%
. Here, the density matrix $\rho _{S}(t)$\ of the system evolves as%
\begin{equation}
\frac{d\rho _{S}(t)}{dt}=\frac{-i}{\hbar }[H_{eff},\rho
_{S}(t)]+\int_{0}^{t}d\tau K(t-\tau )\mathcal{L}[\rho _{S}(\tau )],
\label{memoriosaKernel}
\end{equation}%
where $\mathcal{L}[\bullet ]=-\{D,\bullet \}_{+}+F[\bullet ]$ is a standard
Lindblad superoperator. The memory kernel $K(t)$ is a function that may
introduces strong non-Markovian effects in the system decay dynamics.

The study and characterization of this kind of dynamics is twofold: on one
hand, there is a general fundamental interest in the theory of open quantum
systems to extend the methods and concepts well developed for Markovian
dynamics to the non-Markov case. On the other hand there are many new
physical situations in which the Markov assumption, usually used, is not
fulfill and then non-Markovian dynamics has to be introduced. Remarkable
examples are single fluorescent systems hosted in complex environments \cite%
{barkaiChem,schlegel,brokmann,grigolini,rapid}, superconducting qubits \cite%
{makhlinReport,falci} and band gap materials \cite{john,quang}.

Most of the recent analysis on non-Markovian Lindblad evolutions \cite%
{barnett,wilkie,budini,cresser,lidar,sabrina,maniscalco} were focus on the
possibility of obtaining non-physical solution for $\rho _{S}(t)$ from Eq.~(%
\ref{memoriosaKernel}). This problem was clarified in Refs.~\cite%
{wilkie,budini}, where mathematical constraints on the kernel $K(t)$ that
guarantees the completely positive condition \cite%
{petruccione,alicki,nielsen} of the solution map $\rho _{S}(0)\rightarrow
\rho _{S}(t)$ were found. Furthermore, in Ref.~\cite{budini} the completely
positive condition was associated with the possibility of finding a
stochastic representation of the system dynamics.

There also exist different analysis that associate evolutions like Eq.~(\ref%
{memoriosaKernel}) with microscopic system environment interactions \cite%
{wilkieChem,maniscalco,jpa,gbma}. In Ref.~\cite{jpa} the microscopic
Hamiltonian involves extra stationary unobserved degrees of freedom that
modulate the dissipative coupling between the system of interest an a
Markovian environment. This kind of interaction lead to a Lindblad equation
characterized by a \textit{random rate}. A similar situation was found in
Ref.~\cite{gbma} by considering a complex environment whose action can be
described in a \textit{generalized Born-Markov approximation} (GBMA). This
approach relies in the possibility of splitting the environment in a
\textquotedblleft direct sum\textquotedblright\ of sub-reservoirs, each one
being able to induce by itself a Markovian system evolution. When the
system-environment interaction does not couples the different subspaces
associated to each sub-reservoir, the system dynamics can also be written as
a Lindblad equation with a random dissipative rate. After performing the
average over the random rate, the system dynamics can be written as a
non-local evolution with a structure similar to Eq.~(\ref{memoriosaKernel}).
Besides its theoretical interest, the GBMA was found to be an useful tool
for describing and modeling specific physical situations, such as the
fluorescence signal scattered by individual nanoscopic systems host in
condensed phase environments \cite{rapid}.

The aim of the present work is to go beyond previous results \cite%
{barnett,wilkie,budini,cresser,lidar,sabrina,maniscalco,wilkieChem,jpa,gbma,salo}%
, and present an alternative kind of evolution that induces strong non-local
effects, providing in this way an extra framework for studying and
characterizing non-Markovian open quantum system dynamics. In the present
approach, the system density matrix can be written as%
\begin{equation}
\rho _{S}(t)=\sum_{R}\tilde{\rho}_{R}(t),  \label{Suma}
\end{equation}%
where the unnormalized states $\tilde{\rho}_{R}(t)$ have associated an
effective Hamiltonian $H_{R}^{eff},$ and their full evolution is defined by%
\begin{equation}
\begin{array}{r}
\dfrac{d}{dt}\tilde{\rho}_{R}(t)=\dfrac{-i}{\hbar }[H_{R}^{eff},\tilde{\rho}%
_{R}(t)]-\{D_{R},\tilde{\rho}_{R}(t)\}_{+}+F_{R}[\tilde{\rho}_{R}(t)] \\ 
\\ 
-\sum\limits_{\substack{ R^{\prime }  \\ R^{\prime }\neq R}}\{D_{R^{\prime
}R},\tilde{\rho}_{R}(t)\}_{+}+\sum\limits_{\substack{ R^{\prime }  \\ %
R^{\prime }\neq R}}F_{RR^{\prime }}[\tilde{\rho}_{R^{\prime }}(t)],%
\end{array}
\label{RATE}
\end{equation}%
subject to the initial conditions%
\begin{equation}
\tilde{\rho}_{R}(0)=P_{R}\rho _{S}(0).  \label{Initial}
\end{equation}%
The positive weights $P_{R}$ satisfy $\sum_{R}P_{R}=1.$ On the other hand,
the diagonal superoperator contributions are defined by%
\begin{equation}
D_{R}=\frac{1}{2}\sum_{\alpha ,\gamma }a_{R}^{\alpha \gamma }V_{\gamma
}^{\dagger }V_{\alpha },\ \ \ \ \ F_{R}[\bullet ]=\sum_{\alpha ,\gamma
}a_{R}^{\alpha \gamma }V_{\alpha }\bullet V_{\gamma }^{\dag },
\end{equation}%
while the non-diagonal contributions reads%
\begin{equation}
D_{R^{\prime }R}=\frac{1}{2}\sum_{\alpha ,\gamma }a_{R^{\prime }R}^{\alpha
\gamma }V_{\gamma }^{\dagger }V_{\alpha },\ \ \ \ \ F_{RR^{\prime }}[\bullet
]=\sum_{\alpha ,\gamma }a_{RR^{\prime }}^{\alpha \gamma }V_{\alpha }\bullet
V_{\gamma }^{\dag }.  \label{NonDiagonal}
\end{equation}%
By convenience, we have introduced different notations for the diagonal and
non-diagonal terms. As in standard Lindblad equations, Eq.~(\ref{Lindblad}),
the matrixes $a_{R}^{\alpha \gamma }$ and $a_{R^{\prime }R}^{\alpha \gamma }$
characterize the dissipative rate constants. The structure of the
non-diagonal terms in Eq.~(\ref{RATE}) resemble a classical rate equation 
\cite{kampen}. Therefore, we name this kind of evolution as a \textit{%
Lindblad rate equation. }

Our main objective is to characterize this kind of equations by finding
different microscopic interactions that leads to this structure.
Furthermore, we find the conditions that guarantees that the solution map $%
\rho _{S}(0)\rightarrow \rho _{S}(t)$ is a completely positive one.

While the evolution of $\rho _{S}(t)$ can be written as a non-local
evolution [see Eq.~(\ref{memoria})], the structure Eq.~(\ref{RATE}) leads to
a kind of non-Markovian effects where the stationary properties may depend
on the system initialization. In order to understand this unusual
characteristic, as in Ref.~\cite{budini,gbma}, we also explore the
possibility of finding a stochastic representation of the system dynamics.

We remark that specific evolutions like Eq.~(\ref{RATE}) were derived
previously in the literature in the context of different approaches \cite%
{esposito,breuer,gbma}. The relation between those results is also clarified
in the present contribution.

The paper is organized as follows. Is Sec. II we derive the Lindblad rate
equations from a GBMA by considering interactions Hamiltonians that has
contribution terms between the subspaces associated to each sub-reservoir.
An alternative derivation in terms of tripartite interactions allows to find
the conditions under which the dynamic is completely positive. A third
derivation is given in terms of quantum stochastic processes. In Sec. III we
characterize the resulting non-Markovian master equation. By analyzing some
simple non-trivial examples that admits a stochastic reformulation, we
explain some non-standard general properties of the non-Markovian dynamics.
In Sec. IV we give the conclusions.

\section{Microscopic derivation}

In this section we present three alternative situations where the system
dynamics is described by a Lindblad rate equation.

\subsection{Generalized Born-Markov approximation}

The GBMA applies to complex environments whose action can be well described
in terms of a direct sum of Markovian sub-reservoirs \cite{gbma}. This
hypothesis implies that the total system-environment density matrix, in
contrast with the standard separable form \cite{petruccione,cohen}, assumes
a classical correlated structure \cite{nielsen} (see Eq.~(6) in Ref.~\cite%
{gbma}). In our previous analysis, we have assumed a system-environment
interaction Hamiltonian that does not have matrix elements between the
subspaces associated to each sub-reservoir. Therefore it assumes a direct
sum structure (see Eq.~(5) in Ref.~\cite{gbma}). By raising up this
condition, i.e., by taking in account arbitrary interaction Hamiltonians
without a direct sum structure, it is possible to demonstrate that the GBMA
leads to a Lindblad rate equation, Eq.~(\ref{RATE}).

As in the standard Born-Markov approximation, the derivation of the system
evolution can be formalized in terms of projector techniques \cite{haake}.
In fact, in Ref.~\cite{breuer} Breuer and collaborators introduced a
\textquotedblleft correlated projector technique\textquotedblright\ intended
to describe situations where the total system-environment density matrix
does not assume an uncorrelated structure. Therefore, the system dynamics
can be alternatively derived in the context of this equivalent approach. The
main advantage of this technique is that it provides a rigorous procedure
for obtaining the dynamics to any desired order in the system-environment
interaction strength \cite{haake,breuer}. Here, we assume that the system is
weakly coupled to the environment. Therefore, we work out the system
evolution up to second order in the interaction strength.

We start by considering a full microscopic Hamiltonian description of the
interaction of a system $S$ with its environment $B$%
\begin{equation}
H_{T}=H_{S}+H_{B}+H_{I}.  \label{total}
\end{equation}%
The contributions $H_{S}$ and $H_{B}$ correspond to the system and bath
Hamiltonians respectively. The term $H_{I}$ describes their mutual
interaction.

The system density matrix follows after tracing out the environment degrees
of freedom, $\rho _{S}(t)=\mathrm{Tr}_{B}\{\rho _{T}(t)\},$ where the total
density matrix $\rho _{T}(t)$\ evolves as%
\begin{equation}
\frac{d\rho _{T}(t)}{dt}=\frac{-i}{\hbar }[H_{T},\rho _{T}(t)]\equiv 
\mathcal{L}_{T}[\rho _{T}(t)].  \label{rho}
\end{equation}%
Now, we introduce the projector $\mathcal{P}$ defined by%
\begin{equation}
\mathcal{P}\rho _{T}(t)=\sum_{R}\tilde{\rho}_{R}(t)\otimes \frac{\Xi _{R}}{%
\mathrm{Tr}_{B}\{\Xi _{R}\}},  \label{projector}
\end{equation}%
where $\Xi _{R}$ is given by%
\begin{equation}
\Xi _{R}\equiv \Pi _{R}\rho _{B}\Pi _{R},
\end{equation}%
with $\rho _{B}$ being the stationary state of the bath, while the system
states $\tilde{\rho}_{R}(t)$ are defined by%
\begin{equation}
\tilde{\rho}_{R}(t)\equiv \mathrm{Tr}_{B}\{\Pi _{R}\rho _{T}(t)\Pi _{R}\}.
\end{equation}%
We have introduced a set of projectors $\Pi _{R}=\sum_{\{\epsilon
_{R}\}}|\epsilon _{R}\rangle \langle \epsilon _{R}|,$ which provides an
orthogonal decomposition of the unit operator [$I_{B}$] in the Hilbert space
of the bath, $\sum_{R}\Pi _{R}=I_{B},$ with $\Pi _{R}\Pi _{R^{\prime }}=\Pi
_{R}\delta _{R,R^{\prime }}.$ The full set of states $|\epsilon _{R}\rangle $
corresponds to the base where $\rho _{B}$\ is diagonal, which implies $%
\sum_{R}\Xi _{R}=\rho _{B}.$

It is easy to realize that $\mathcal{P}^{2}=\mathcal{P}.$ In physical terms,
this projector takes in account that each bath-subspace associated to the
projectors $\Pi _{R}$ induces a different system dynamics, each one
represented by the states $\tilde{\rho}_{R}(t).$ Each sub-space can be seen
as a sub-reservoir. On the other hand, notice that the standard projector $%
\mathcal{P}\rho _{T}(t)=\mathrm{Tr}_{B}\{\rho _{T}(t)\}\otimes \rho
_{B}=\rho _{S}(t)\otimes \rho _{B}$ \cite{haake}, is recuperated when all
the states $\tilde{\rho}_{R}(t)$ have the same dynamics. Therefore, it is
evident that the definition of the projector Eq.~(\ref{projector}) implies
the introduction of a \textit{generalized Born approximation }\cite{gbma},
where instead of a uncorrelated form for the total system-environment
density matrix, it is assumed a classical correlated state.

By using that $\sum_{R}\Pi _{R}=I_{B},$ the system density matrix can be
written as 
\begin{subequations}
\label{sum}
\begin{eqnarray}
\rho _{S}(t) &=&\sum_{R}\mathrm{Tr}_{B}\{\Pi _{R}\rho _{T}(t)\Pi _{R}\}\frac{%
\mathrm{Tr}_{B}\{\Xi _{R}\}}{\mathrm{Tr}_{B}\{\Xi _{R}\}} \\
&=&\mathrm{Tr}_{B}\{\mathcal{P}\rho _{T}(t)\}=\sum_{R}\tilde{\rho}_{R}(t)
\end{eqnarray}
This equation defines the system state as a sum over the states $\tilde{\rho}%
_{R}(t).$ Notice that the second line follows from the definition of the
objects that define the projector Eq.~(\ref{projector}).

By writing the evolution Eq.~(\ref{rho}) in an interaction representation
with respect to $H_{S}+H_{B},$ and splitting the full dynamics in the
contributions $\mathcal{P}\rho _{T}(t)$ and $\mathcal{Q}\rho _{T}(t),$ where 
$\mathcal{Q}=1-\mathcal{P},$ up to second order in the interaction
Hamiltonian it follows \cite{haake} 
\end{subequations}
\begin{equation}
\frac{d\mathcal{P}\rho _{T}(t)}{dt}=\int_{0}^{t}dt^{\prime }\mathcal{PL}%
_{T}(t)\mathcal{L}_{T}(t^{\prime })\mathcal{P}\rho _{T}(t^{\prime }),
\label{Interaction}
\end{equation}%
where $\mathcal{L}_{T}(t)$ is the total Liouville superoperator in a
interaction representation. For writing the previous equation, we have
assumed $\mathcal{Q}\rho _{T}(0)=0,$ which implies the absence of any\
initial correlation between the system and the bath, $\rho _{T}(0)=\rho
_{S}(0)\otimes \rho _{B}.$ Then, the initial condition of each state $\tilde{%
\rho}_{R}(t)$ can be written as%
\begin{equation}
\tilde{\rho}_{R}(0)=P_{R}\rho _{S}(0).
\end{equation}%
The parameters $P_{R}$ are defined by the weight of each sub-reservoir in
the full stationary bath state%
\begin{equation}
P_{R}=\mathrm{Tr}_{B}\{\Xi _{R}\}=\mathrm{Tr}_{B}\{\Pi _{R}\rho
_{B}\}=\sum_{\{\epsilon _{R}\}}\langle \epsilon _{R}|\rho _{B}|\epsilon
_{R}\rangle ,  \label{pesos}
\end{equation}%
which trivially satisfies $\sum_{R}P_{R}=1.$

Now, we split the interaction Hamiltonian as%
\begin{equation}
H_{I}=\sum_{R,R^{\prime }}H_{I_{RR^{\prime }}}\equiv \sum_{R,R^{\prime }}\Pi
_{R}H_{I}\Pi _{R^{\prime }}.  \label{HI}
\end{equation}
We notice that when $\Pi _{R}H_{I}\Pi _{R^{\prime }}=0$ for $R\neq R^{\prime
},$ the interaction Hamiltonian can be written as a direct sum $%
H_{I}=H_{I_{1}}\oplus H_{I_{2}}\cdots \oplus H_{I_{R}}\oplus
H_{I_{R+1}}\cdots ,$ with $H_{I_{R}}=\Pi _{R}H_{I}\Pi _{R}.$ This case
recover the assumptions made in Ref. \cite{gbma}. In fact, without
considering the non-diagonal terms in Eq.~(\ref{RATE}) $[a_{RR^{\prime
}}^{\alpha \gamma }=0],$ after a trivial change of notation $\tilde{\rho}%
_{R}(t)\rightarrow P_{R}\rho _{R}(t)$ in Eq.~(\ref{Suma}), the dynamics
reduce to a random Lindblad equation.

In order to proceed with the present derivation, we introduce the
superoperator identity \cite{bariloche}%
\begin{equation}
\lbrack \hat{a},[\hat{b},\bullet ]]=\frac{1}{2}[[\hat{a},\hat{b}],\bullet ]+%
\frac{1}{2}\{\{\hat{a},\hat{b}\}_{+},\bullet \}_{+}-(\hat{a}\bullet \hat{b}+%
\hat{b}\bullet \hat{a}),
\end{equation}%
valid for arbitrary operators $\hat{a}$ and $\hat{b}.$ By using this
identity and the splitting Eq.~(\ref{HI}) into Eq.~(\ref{Interaction}),
after a straightforward calculation the evolution of $\tilde{\rho}_{R}(t)$
in the Schr\"{o}dinger representation can be written as in Eq.~(\ref{RATE}).
The effective Hamiltonians read%
\begin{equation}
H_{R}^{eff}=H_{S}-i\frac{\hbar }{2}\int_{0}^{\infty }d\tau \mathrm{Tr}%
_{B_{R}}\{[H_{I},H_{I}(-\tau )]\rho _{B_{R}}\}.  \label{Hefectivo}
\end{equation}%
The non-diagonal operators $D_{R^{\prime }R}$ read 
\begin{equation}
D_{R^{\prime }R}=\frac{1}{2}\int_{0}^{\infty }d\tau \mathrm{Tr}%
_{B_{R}}([H_{I_{RR^{\prime }}}H_{I_{R^{\prime }R}}(-\tau )+h.c.]\rho
_{B_{R}}),  \label{Drr}
\end{equation}%
while the corresponding superoperators $F_{RR^{\prime }}$ can be written as%
\begin{equation}
F_{RR^{\prime }}[\bullet ]=\int_{0}^{\infty }d\tau \mathrm{Tr}%
_{B_{R}}(H_{I_{RR^{\prime }}}(-\tau )[\bullet ]\otimes \rho _{B_{R^{\prime
}}}H_{I_{R^{\prime }R}}+h.c.).  \label{Frr}
\end{equation}%
The diagonal contributions follows from the previous expressions as $%
D_{R}=D_{RR},$ and $F_{R}[\bullet ]=F_{RR}[\bullet ].$ Furthermore, we have
defined $\mathrm{Tr}_{B_{R}}\{\bullet \}\equiv \mathrm{Tr}_{B}\{\Pi
_{R}\bullet \Pi _{R}\}$ and%
\begin{equation}
\rho _{B_{R}}\equiv \Xi _{R}/P_{R}.  \label{RhoSubBath}
\end{equation}%
Notice that these objects correspond to the stationary state of each
sub-reservoir.

In obtaining Eqs.~(\ref{Hefectivo})\ to (\ref{Frr}) we have introduced a
standard \textit{Markovian approximation }\cite{petruccione,cohen}, which
allows to obtain local in time evolutions for the set $\{\tilde{\rho}%
_{R}(t)\},$ as well as to extend the time integrals to infinite. This
approximation applies when the diagonal and non-diagonal correlations of the
different sub-reservoirs define the small time scale of the problem. In
order to clarify the introduction of the Markov approximation, we assume
that the interaction Hamiltonian can be written as%
\begin{equation}
H_{I}=\sum_{\alpha }V_{\alpha }\otimes B_{\alpha },
\end{equation}%
where the operators $V_{\alpha }$ and $B_{\alpha }$ act on the system and
bath Hilbert spaces respectively. By using $H_{I}=H_{I}^{\dag },$ the
previous expressions Eqs.~(\ref{Drr}) and (\ref{Frr}) read%
\begin{equation}
D_{R^{\prime }R}=\frac{1}{2}\sum_{\alpha \beta }\int_{0}^{\infty }d\tau
\{\chi _{R^{\prime }R}^{\alpha \beta }(-\tau )V_{\alpha }^{\dagger }V_{\beta
}(-\tau )+h.c.\},  \label{DrrBis}
\end{equation}%
and%
\begin{equation}
F_{RR^{\prime }}[\bullet ]=\sum_{\alpha \beta }\int_{0}^{\infty }\{d\tau
\chi _{RR^{\prime }}^{\alpha \beta }(-\tau )V_{\beta }(-\tau )[\bullet
]V_{\alpha }^{\dagger }+h.c.\}.  \label{FrrBis}
\end{equation}%
Here, we have defined the \textquotedblleft \textit{projected bath
correlations}\textquotedblright 
\begin{equation}
\chi _{RR^{\prime }}^{\alpha \beta }(-\tau )\equiv \mathrm{Tr}_{B_{R^{\prime
}}}\{\rho _{B_{R^{\prime }}}B_{\alpha }^{\dagger }\Pi _{R}B_{\beta }(-\tau
)\}.  \label{ProjectedCorrelation}
\end{equation}%
Without taking in account the indexes $R$ and $R^{\prime },$ this expression
reduces to the standard definition of bath correlation \cite%
{petruccione,cohen,alicki,bariloche}. Here, the same structure arises with
projected elements. As the integrals that appears in Eqs.~(\ref{DrrBis}) and
(\ref{FrrBis}) have the same structure that in the standard Born-Markov
approximation \cite{bariloche}, the meaning of the previous calculation
steps becomes clear.

Finally, in order to obtain the explicit expressions for the matrixes $%
a_{RR^{\prime }}^{\alpha \gamma }$ and $a_{R}^{\alpha \gamma },$ we define a
matrix $C_{\beta \gamma }(-\tau )$ from%
\begin{equation}
V_{\beta }(-\tau )=e^{-i\tau H_{S}}V_{\beta }e^{+i\tau
H_{S}}=\sum\limits_{\gamma }C_{\beta \gamma }(-\tau )V_{\gamma }.
\end{equation}%
By introducing these coefficients in Eqs.~(\ref{DrrBis}) and (\ref{FrrBis}),
it is possible to write the operators $D_{R^{\prime }R}$ and $F_{RR^{\prime
}}[\bullet ]$ as in Eq.~(\ref{NonDiagonal}). The matrix $a_{RR^{\prime
}}^{\alpha \gamma }$ is defined by%
\begin{eqnarray}
a_{RR^{\prime }}^{\alpha \gamma } &=&\sum_{\beta }\int_{0}^{\infty }d\tau
\chi _{RR^{\prime }}^{\gamma \beta }(-\tau )\ C_{\beta \alpha }(-\tau ) 
\notag \\
&&+\sum_{\beta }\int_{0}^{\infty }d\tau (\chi _{RR^{\prime }}^{\alpha \beta
})^{^{\ast }}\!(-\tau )\ C_{\beta \gamma }^{\ast }(-\tau ),  \label{aRR}
\end{eqnarray}%
while the diagonal matrix elements follows as $a_{R}^{\alpha \gamma
}=a_{RR}^{\alpha \gamma }.$ Consistently, without taking in account the
indexes $R$ and $R^{\prime },$ this matrix structure reduce to that of the
standard Born-Markov approximation \cite{bariloche}.

\subsubsection*{Quantum master equation for a system influencing its
environment}

In Ref.~\cite{esposito}, Esposito and Gaspard deduced a quantum master
equation intended to describe physical situations where the density of
states of a reservoir is affected by the changes of energy of an open
system. While this physical motivation is different to that of the GBMA \cite%
{gbma} (or in general, to the correlated projector techniques \cite{breuer}%
), here we show that both formalisms can be deduced by using the same
calculations steps. Therefore, the evolution of Ref.~\cite{esposito} can
also be written as a Lindblad rate equation.

In Ref.~\cite{esposito}, the system evolution is derived by taking in
account the effect of the energy exchanges between the system and the
environment and the conservation of energy by the total (closed)
system-reservoir dynamics. These conditions are preserved by tracing-out the
bath coherences and maintaining all the information with respect to the bath
populations. Therefore, the system density matrix is written in terms of an
auxiliary state that depends parametrically on the energy\ of the
environment, which is assumed in a microcanonical state. By noting that in
the GBMA there not exist any coherence between the different sub-reservoirs
[see Eq.~(\ref{projector})], we realize that the dynamics obtained in Ref.~%
\cite{esposito} can be recovered with the previous results by associating
the \textit{discrete} index $R$ with a \textit{continuos} parameter $%
\varepsilon ,$ which label the eigenvalues of the reservoir, joint with the
replacements%
\begin{equation}
\tilde{\rho}_{R}(t)\rightarrow \tilde{\rho}(\varepsilon ;t),\ \ \ \ \ \ \ \
\ \ \ \sum_{R}\rightarrow \int d\varepsilon \ n(\varepsilon ),
\end{equation}%
where $n(\varepsilon )$\ is the spectral density function of the reservoir.
Consistently, the system state [Eq.~(\ref{sum})] is written as%
\begin{equation}
\rho _{S}(t)=\int d\varepsilon \ n(\varepsilon )\tilde{\rho}(\varepsilon
;t)\equiv \int d\varepsilon \ \bar{\rho}(\varepsilon ;t).
\end{equation}%
As in the GBMA, the evolution of $\bar{\rho}(\varepsilon ;t)$ can be written
as a Lindblad rate equation defined in terms of the matrix structure Eq.~(%
\ref{aRR}) with the replacement $\chi _{RR^{\prime }}^{\alpha \beta }(-\tau
)\rightarrow \chi _{\varepsilon \varepsilon ^{\prime }}^{\alpha \beta
}(-\tau ),$ where%
\begin{equation}
\chi _{\varepsilon \varepsilon ^{\prime }}^{\alpha \beta }(-\tau
)=\left\langle \varepsilon ^{\prime }\right\vert B_{\alpha }^{\dagger
}\left\vert \varepsilon \right\rangle \left\langle \varepsilon \right\vert
B_{\beta }\left\vert \varepsilon ^{\prime }\right\rangle \exp
[-i(\varepsilon -\varepsilon ^{\prime })\tau ].
\end{equation}%
This last definition follows from the microcanonical state of the reservoir $%
[\rho _{B}\rightarrow 1].$ Finally, by introducing the matrix elements%
\begin{equation}
P_{ss^{\prime }}(\varepsilon ;t)\equiv \langle s|\bar{\rho}(\varepsilon
;t)|s^{\prime }\rangle ,
\end{equation}%
where $\{|s\rangle \}$ are the eigenstates of the system Hamiltonian, $%
H_{S}|s\rangle =\varepsilon _{s}|s\rangle ,$ the master equation of Ref.~%
\cite{esposito} is explicitly recovered. Due to the energy preservation
condition, in general the evolution involves a continuos parametric coupling
between the matrix elements $P_{ss^{\prime }}(\varepsilon ;t)$ and $%
P_{ss^{\prime }}(\varepsilon \pm \Delta ;t),$ where $\Delta $\ is a energy
scale that characterize the natural transition frequencies of the system 
\cite{esposito}.

We remark that the difference between both approaches relies on the assumed
properties of the environment. In the context of the GBMA, the index $R$
label a set of Hilbert subspaces each one defined in terms of a manifold of
bath eigenstates able to induce, by itself, a Markovian system dynamics.
Therefore, by hypothesis, the complete environment does not feels the
effects of the system energy changes. On the other hand, the approach of
Esposito and Gaspard applies to the opposite situation where, by hypothesis,
the density of states of the environment vary on a scale comparable to the
system energy transitions. The stretched similarity between both approaches
follows from the absence of coherences between the different (discrete or
continuous) bath sub-spaces. In both cases the system evolution can be
written as a Lindblad rate equation.

\subsection{Composite environments}

The previous analysis relies in a bipartite system-environment interaction
described in a GBMA. Here, we arrive to a Lindblad rate equation by
considering composite environments, where extra degrees of freedom $U$
modulate the interaction (the entanglement) between a system $S$ and a
Markovian reservoir $B$ \cite{jpa}. This formulation allows to find the
conditions under which Eq.~(\ref{RATE}) defines a completely positive
evolution.

The total Hamiltonian reads%
\begin{equation}
H_{T}=H_{S}+H_{U}+H_{SU}+H_{B}+H_{I}.
\end{equation}
As before, $H_{S}$ represent the system Hamiltonian. Here, $H_{B}$ is the
Hamiltonian of the Markovian environment. On the other hand, $H_{U}$ is the
Hamiltonian of the extra degrees of freedom that modulate the
system-environment interaction. The interaction Hamiltonian $H_{I}$ couples
the three involved parts. We also consider the possibility of a direct
interaction between $S$ and $U,$ denoted by $H_{SU}.$

As $B$ is a Markovian reservoir, we can trace out its degrees of freedom in
a standard way \cite{petruccione,cohen,alicki}. Therefore, we assume the
completely positive Lindblad evolution%
\begin{equation}
\frac{d\rho _{C}(t)}{dt}=\frac{-i}{\hbar }[H_{C},\rho _{C}(t)]-\{D_{C},\rho
_{C}(t)\}_{+}+F_{C}[\rho _{C}(t)],  \label{RhoC}
\end{equation}
with the definitions 
\begin{equation}
D_{C}=\frac{1}{2}\sum_{i,j}b_{ij}A_{j}^{\dagger
}A_{i},\;\;\;\;\;\;\;F_{C}[\bullet ]=\sum_{i,j}b_{ij}A_{i}\bullet
A_{j}^{\dagger }.  \label{DFTripartito}
\end{equation}%
The matrix $\rho _{C}(t)$ corresponds to the state of \ the
\textquotedblleft compose system\textquotedblright\ $S$-$U$ with Hilbert
space $\mathcal{H}_{C}=\mathcal{H}_{S}\otimes \mathcal{H}_{U}.$ The sum
indexes $i$ and $j$ run from one to 1 to $(\dim \mathcal{H}_{C})^{2},$ with $%
\dim \mathcal{H}_{C}=\dim \mathcal{H}_{S}\dim \mathcal{H}_{U}.$
Consistently, the set $\{A_{i}\}$ is a base of operators in $\mathcal{H}%
_{C}, $ and $b_{ij}$ is an arbitrary Hermitian semipositive matrix.

In order to get the system state it is also necessary to trace out the
degrees of freedom $U.$ In fact, $\rho _{S}(t)=\mathrm{Tr}_{U}\{\rho
_{C}(t)\},$ which deliver 
\begin{eqnarray}
\rho _{S}(t) &=&\mathrm{Tr}_{U}\{\rho _{C}(t)\}=\sum_{R}\left\langle
R\right\vert \rho _{C}(t)\left\vert R\right\rangle ,  \notag \\
&\equiv &\sum_{R}\tilde{\rho}_{R}(t).  \label{suma}
\end{eqnarray}
where $\{\left\vert R\right\rangle \}$ is a base of vector states in $%
\mathcal{H}_{U}.$ We notice that here, the sum structure Eq.~(\ref{Suma})
have a trivial interpretation in terms of a trace operation.

By assuming an uncorrelated initial condition $\rho _{C}(0)=\rho
_{S}(0)\otimes \rho _{U}(0),$ where $\rho _{S}(0)$ and $\rho _{U}(0)$ are
arbitrary initial states for the systems $S$ and $U,$ from Eq.~(\ref{suma})
it follows the initial conditions $\tilde{\rho}_{R}(0)=P_{R}\rho _{S}(0),$
where%
\begin{equation}
P_{R}=\left\langle R\right\vert \rho _{U}(0)\left\vert R\right\rangle .
\label{peso}
\end{equation}
Therefore, here the weights $P_{R}$ corresponding to Eq.~(\ref{Initial}) are
defined by the diagonal matrix elements of the initial state of the system $%
U.$ From now on, we will assume that the set of states $\{\left\vert
R\right\rangle \}$ correspond to the eigenvectors basis of $H_{U},$ i.e.,%
\begin{equation}
H_{U}\left\vert R\right\rangle =\varepsilon _{R}\left\vert R\right\rangle .
\label{Eigen}
\end{equation}

The evolution of the states $\tilde{\rho}_{R}(t)=\left\langle R\right\vert
\rho _{C}(t)\left\vert R\right\rangle $ can be obtained from Eq.~(\ref{RhoC}%
) after tracing over system $U.$ Under special \textit{symmetry conditions},
the resulting evolution can be cast in the form of a Lindblad rate equation,
Eq.~(\ref{RATE}). In fact, in a general case, there will be extra
contributions proportional to the components $\left\langle R\right\vert \rho
_{C}(t)\left\vert R^{\prime }\right\rangle .$ By noting that%
\begin{equation}
\mathrm{Tr}_{S}[\left\langle R\right\vert \rho _{C}(t)\left\vert R^{\prime
}\right\rangle ]=\left\langle R\right\vert \rho _{U}(t)\left\vert R^{\prime
}\right\rangle ,
\end{equation}
where $\rho _{U}(t)=\mathrm{Tr}_{S}\{\rho _{C}(t)\}$ is the density matrix
of the degrees of freedom $U,$ we realize that the evolution of $\tilde{\rho}%
_{R}(t)$ can be written as a Lindblad rate equation only when the evolution
of $\rho _{U}(t)$ does not involves coupling between the populations $%
\left\langle R\right\vert \rho _{U}(t)\left\vert R\right\rangle $ and
coherences $\left\langle R\right\vert \rho _{U}(t)\left\vert R^{\prime
}\right\rangle ,$ $R\neq R^{\prime },$ of system $U.$ As is well known \cite%
{petruccione,cohen,alicki}, this property is satisfied when the dissipative
evolution of $\rho _{U}(t)$ can be written in terms of the eigenoperators $%
L_{u}$\ of the unitary dynamic, i.e., $[H_{U},L_{u}]=\omega _{u}L_{u}.$ In
what follows, we show explicitly that this property is \textit{sufficient}
to obtain a Lindblad rate equation for the set of matrixes $\{\tilde{\rho}%
_{R}(t)\}.$

First, we notice that the Hamiltonian $H_{C}$ in Eq.~(\ref{RhoC}) must to
have the structure 
\begin{equation}
H_{C}=H_{S}+H_{U}+\sum_{\alpha }V_{\alpha }\otimes L_{0}^{\alpha },
\end{equation}%
where $L_{0}^{\alpha }$ are the eigenoperators with a null eigenvalue, i.e., 
$[H_{U},L_{0}^{\alpha }]=0.$ With this structure, the populations and
coherences corresponding to $U$ do not couple between them. Therefore, the
effective Hamiltonian $H_{R}^{eff}$in Eq.~(\ref{RATE}) reads%
\begin{equation}
H_{R}^{eff}=H_{S}+\sum_{\alpha }\left\langle R\right\vert L_{0}^{\alpha
}\left\vert R\right\rangle V_{\alpha }.
\end{equation}

After taking the operator base in $\mathcal{H}_{C}=\mathcal{H}_{S}\otimes 
\mathcal{H}_{U}$ as 
\begin{equation}
\{A_{i}\}\rightarrow \{V_{\alpha }\otimes L_{u}\},
\end{equation}%
the superoperators Eq.~(\ref{DFTripartito}) can be written as 
\begin{subequations}
\label{Compuesto}
\begin{eqnarray}
D_{C} &=&\frac{1}{2}\sum_{\substack{ \alpha ,\gamma  \\ u,v}}b_{uv}^{\alpha
\gamma }V_{\gamma }^{\dagger }L_{v}^{\dagger }V_{\alpha }L_{u}, \\
F_{C}[\bullet ] &=&\sum_{\substack{ \alpha ,\gamma  \\ u,v}}b_{uv}^{\alpha
\gamma }V_{\alpha }L_{u}\bullet V_{\gamma }^{\dagger }L_{v}^{\dagger }.
\end{eqnarray}%
With these definitions, by taking the trace operation over the system $U$ in
the evolution Eq.~(\ref{RhoC}), we notice that the evolution of the set $\{%
\tilde{\rho}_{R}(t)\}$ can be cast in the form of a Lindblad rate equation
if the conditions 
\end{subequations}
\begin{equation}
\sum_{u,v}b_{uv}^{\alpha \gamma }\left\langle R^{\prime \prime }\right\vert
L_{v}^{\dag }\left\vert R\right\rangle \left\langle R\right\vert
L_{u}\left\vert R^{\prime }\right\rangle =\delta _{R^{\prime },R^{\prime
\prime }}\,a_{RR^{\prime }}^{\alpha \gamma }  \label{Condiciones}
\end{equation}%
are satisfied. The factor $\delta _{R^{\prime },R^{\prime \prime }}$
guarantees that the evolution of the set $\{\tilde{\rho}_{R}(t)\}$ do not
involve the terms $\left\langle R\right\vert \rho _{C}(t)\left\vert
R^{\prime }\right\rangle ,$ $R\neq R^{\prime },$ and in turn implies that
the populations and coherences of $U$ do not couple between them. On the
other hand, $a_{RR^{\prime }}^{\alpha \gamma }$ defines the matrix elements
corresponding to the structure Eq.~(\ref{RATE}). The diagonal contributions
follows from Eq.~(\ref{Condiciones}) by taking $R=R^{\prime }.$

The set of conditions Eq.~(\ref{Condiciones}) can be simplified by taking
the base%
\begin{equation}
L_{u}\rightarrow |\mathcal{R}^{\prime }\rangle \!\langle \mathcal{R}|,
\end{equation}%
which from Eq.~(\ref{Eigen}) satisfy $[H_{U},L_{u}]=(\varepsilon _{\mathcal{R%
}}-\varepsilon _{\mathcal{R}^{\prime }})L_{u}.$ Thus, Eq.~(\ref{Condiciones}%
) can be consistently satisfied if we impose%
\begin{equation}
b_{uv}^{\alpha \gamma }=0,\ \ \ \ \ \ for\ \ \ \ \ \ u\neq v.
\end{equation}%
After changing $\sum_{u}\!\rightarrow $\negthinspace $\sum_{\mathcal{R,R}%
^{\prime }}$ in Eq.~(\ref{Condiciones}), we get%
\begin{equation}
a_{RR^{\prime }}^{\alpha \gamma }=b_{(R,R^{\prime })(R,R^{\prime })}^{\alpha
\gamma },\ \ \ \ \ a_{R}^{\alpha \gamma }=b_{(R,R)(R,R)}^{\alpha \gamma },
\label{define}
\end{equation}%
where we have used that $\mathcal{R}$ and $\mathcal{R}^{\prime }$ are dumb
indexes. This result demonstrate that the evolution induced by the composite
environment can in fact be written as a Lindblad rate evolution Eq.~(\ref%
{RATE}) with the matrix elements defined by Eq.~(\ref{define}).

From our previous considerations we deduce that Lindblad rate equation arise
from microscopic tripartite interactions having the structure%
\begin{equation}
H_{I}=L_{0}\otimes H_{SB}+\sum_{u}L_{u}\otimes H_{SB}^{u}+L_{u}^{\dag
}\otimes (H_{SB}^{u})^{\dag },  \label{trilogia}
\end{equation}%
where $[H_{U},L_{0}]=0,$ and $L_{u}\rightarrow |R\rangle \!\langle R^{\prime
}|$ with $R\neq R^{\prime }.$ On the other hand, $H_{SB}^{u}$ are arbitrary
interaction terms between the system $S$ and the Markovian environment $B.$
In fact, the structure Eq.~(\ref{trilogia}) guarantees that the populations
and coherences of $U$ do not couple between them, which in turn implies that
the evolutions of the system $S$ is given by a Lindblad rate equation.

\subsubsection*{Completely positive condition}

We have presented two different microscopic interactions that lead to a
Lindblad rate equation. In order to use these equations as a valid tool for
modeling open quantum system dynamics it is necessary to establish the
conditions under which the solution map $\rho _{S}(0)\rightarrow \rho
_{S}(t) $ is a completely positive one. For an arbitrary Lindblad rate
equation this condition must to be defined in terms of the matrixes $%
a_{RR^{\prime }}^{\alpha \gamma }$ and $a_{R}^{\alpha \gamma }.$

In order to find the allowed matrix structures, we notice that the evolution
Eq.~(\ref{RhoC}) is a completely positive one when $b_{ij}\rightarrow
b_{(R,R^{\prime })(R,R^{\prime })}^{\alpha \gamma }$ is a semipositive
defined matrix. Therefore, by using Eq.~(\ref{define}) we arrive to the
conditions%
\begin{equation}
|a_{RR^{\prime }}^{\alpha \gamma }|\geq 0,\ \ \ \ \ \ \ |a_{R}^{\alpha
\gamma }|\geq 0,\ \ \ \ \ \ \ \ \forall \,R,R^{\prime },
\label{CondicionesPositividad}
\end{equation}
i.e., for any value of $R$ and $R^{\prime }$ both kind of matrixes must to
be semipositive defined in the system indexes $\alpha ,\gamma $. The
condition $|a_{R}^{\alpha \gamma }|\geq 0$ has a trivial interpretation. In
fact, when $a_{RR^{\prime }}^{\alpha \gamma }=0,$ there not exist any
dynamical coupling between the states $\tilde{\rho}_{R}(t).$ Thus, their
evolutions are defined by a Lindblad structure that under the constraint $%
|a_{R}^{\alpha \gamma }|\geq 0$ define a completely positive evolution.

\subsection{Quantum random walk}

By using the similarity of Eq.~(\ref{RATE}) with a classical rate equation 
\cite{kampen}, here we present a third derivation by constructing a
stochastic dynamics that develops in the system Hilbert space and whose
average evolution is given by a Lindblad rate equation.

First, we assume that the system is endowed with a classical internal degree
of freedom characterized by a set $\{R\}$ of possible states. The
corresponding populations $P_{R}(t)$ obey the classical evolution%
\begin{equation}
\frac{dP_{R}(t)}{dt}-\sum\limits_{\substack{ R^{\prime }  \\ R^{\prime }\neq
R }}\gamma _{R^{\prime }R}P_{R}(t)+\sum\limits_{\substack{ R^{\prime }  \\ %
R^{\prime }\neq R}}\gamma _{RR^{\prime }}P_{R^{\prime }}(t),
\end{equation}%
with initial conditions $P_{R}(0)=P_{R},$ and where the coefficients $%
\{\gamma _{R^{\prime }R}\}$\ define the hopping rates between the different
classical states $R.$

To each state $R$ we associate a different Markovian system dynamics, whose
evolution is generated by the superoperator%
\begin{equation}
\mathcal{\bar{L}}_{R}=\mathcal{L}_{H}+\mathcal{L}_{R},  \label{self}
\end{equation}
with $\mathcal{L}_{H}[\bullet ]=(-i/\hbar )[H_{S},\bullet ]$ and a standard
Lindblad contribution $\mathcal{L}_{R}[\bullet ]=-\{D_{R},\bullet
\}_{+}+F_{R}[\bullet ].$ Therefore, each state $R$ defines a \textit{\
propagation channel} with a different self-dynamic. The system state follows
by tracing out any information about the internal state. Thus, we write 
\begin{equation}
\rho _{S}(t)=\sum_{R}\tilde{\rho}_{R}(t),
\end{equation}
where each state $\tilde{\rho}_{R}(t)$ defines the system state \textit{given%
} that the internal degree of freedom is in the state $R.$ Consistently, the
initial condition of the auxiliary states reads $\tilde{\rho}%
_{R}(0)=P_{R}\rho _{S}(0).$

Finally, we assume that in each transition $R\rightarrow R^{\prime }$ of the
internal degree of freedom, it is applied a completely positive
superoperator $\mathcal{E}_{R}$ \cite{petruccione,alicki,nielsen}, which
produces a disruptive transformation in the system state.

The stochastic dynamics is completely defined after providing the
self-channel dynamics, defined by $\{\mathcal{\bar{L}}_{R}\},$ the\ set of
rates $\{\gamma _{R^{\prime }R}\}$ and the superoperators $\{\mathcal{E}%
_{R}\}.$ By construction this dynamics is completely positive. The explicit
construction of the corresponding stochastic realizations, which develop in
the system Hilbert space, is as follows. When the system is effectively in
channel $R,$ it is transferred to channel $R^{\prime }$ with rate $\gamma
_{R^{\prime }R}.$ Therefore, the probability of staying in channel $R$
during a sojourn interval $t$ is given by%
\begin{equation}
P_{0}^{(R)}(t)=\exp [-t\dsum\limits_{\substack{ R^{\prime }  \\ R^{\prime
}\neq R}}\gamma _{R^{\prime }R}].
\end{equation}%
This function completely defines the statistics of the time intervals
between the successive disruptive events. As in standard classical rate
equations, when the system \textquotedblleft jump outside\textquotedblright\
of channel $R,$ each subsequent channel $R^{\prime }$ is selected with
probability 
\begin{equation}
t_{R^{\prime }R}=\frac{\gamma _{R^{\prime }R}}{\sum_{\substack{ R^{\prime
\prime }  \\ R^{\prime \prime }\neq R}}\gamma _{R^{\prime \prime }R}},
\end{equation}%
in such a way that $\sum_{R^{\prime }}t_{R^{\prime }R}=1.$ Furthermore, each
transference $R\rightarrow R^{\prime },$ is attended by the application of
the superoperator $\mathcal{E}_{R},$ which produces the disruptive
transformation $\tilde{\rho}_{R}(t)\rightarrow \mathcal{E}_{R}[\tilde{\rho}%
_{R}(t)].$ This transformed state is the subsequent initial condition for
channel $R^{\prime }.$

The average over realizations of the previous quantum stochastic process,
for each state $\tilde{\rho}_{R}(t),$ reads%
\begin{eqnarray}
\tilde{\rho}_{R}(t) &=&P_{0}^{(R)}(t)e_{R}^{t\mathcal{\bar{L}}_{R}}\tilde{%
\rho}(0)+\dint_{0}^{t}d\tau P_{0}^{(R)}(t-\tau )e^{(t-\tau )\mathcal{\bar{L}}%
_{R}}  \notag \\
&&\times \dsum\limits_{\substack{ R^{\prime }  \\ R^{\prime }\neq R}}\gamma
_{RR^{\prime }}\mathcal{E}_{R^{\prime }}[\tilde{\rho}_{R^{\prime }}(\tau )],
\label{facil}
\end{eqnarray}
The structure of this equation has a clear interpretation. The first
contribution represents the realization where the system remains in channel $%
R$ without happening any scattering event. Clearly this term must be
weighted by the probability of not having any event in the time interval $%
(t,0),$ i.e., with the probability $P_{0}^{(R)}(t).$ On the other hand, the
terms inside the integral correspond to the rest of the realizations. They
take in account the contributions that come from any other channel $%
R^{\prime },$ arriving at time $\tau $ and surviving up to time $t$ in
channel $R.$ During this interval it is applied the self-channel propagator $%
\exp [(t-\tau )\mathcal{\bar{L}}_{R}].$ As before, this evolution is
weighted by the survival probability $P_{0}^{(R)}(t-\tau ).$

By working Eq.~(\ref{facil}) in the Laplace domain, after a simple
calculation, it is possible to arrive to the evolution%
\begin{equation}
\begin{array}{r}
\dfrac{d}{dt}\tilde{\rho}_{R}(t)=\dfrac{-i}{\hbar }[H_{S},\tilde{\rho}%
_{R}(t)]-\{D_{R},\tilde{\rho}_{R}(t)\}_{+}+F_{R}[\tilde{\rho}_{R}(t)] \\ 
\\ 
-\sum\limits_{\substack{ R^{\prime }  \\ R^{\prime }\neq R}}\gamma
_{R^{\prime }R}\tilde{\rho}_{R}(t)+\sum\limits_{\substack{ R^{\prime }  \\ %
R^{\prime }\neq R}}\gamma _{RR^{\prime }}\mathcal{E}_{R^{\prime }}[\tilde{
\rho}_{R^{\prime }}(t)].%
\end{array}
\label{Walk}
\end{equation}
We notice that this expression does not corresponds to the more general
structure of a Lindblad rate equation, Eq.~(\ref{RATE}). Nevertheless, there
exist different non-trivial situations that fall in this category. As we
demonstrate in the next section, the advantage of this formulation is that
it provides a simple framework for understanding some non-usual
characteristics of the system dynamics.

\section{Non-Markovian dynamics}

In this section we obtain the master equation that define the evolution of
the system state $\rho _{S}(t)$ associated to an arbitrary Lindblad rate
equation, Eq.~(\ref{RATE}).

In order to simplify the notation, we define a column vector\ defined in the 
$R$-space and whose elements are the states $\tilde{\rho}_{R},$ i.e., $%
\left\vert \tilde{\rho}\right) =(\tilde{\rho}_{1},\tilde{\rho}_{2},\ldots 
\tilde{\rho}_{R},\ldots )^{\mathrm{T}},$ where\ $\mathrm{T}$ denote a
transposition operation. Then, the evolution Eq.~(\ref{RATE}) can be written
as%
\begin{equation}
\frac{d\left\vert \tilde{\rho}(t)\right) }{dt}=\mathcal{L}_{H}\left\vert 
\tilde{\rho}(t)\right) +\mathbb{\hat{M}}\left\vert \tilde{\rho}(t)\right) .
\label{local}
\end{equation}%
where $\mathcal{L}_{H}[\bullet ]=-(i/\hbar )[H_{S},\bullet ],$ and the
matrix elements of $\mathbb{\hat{M}}$\ reads%
\begin{eqnarray}
\mathbb{\hat{M}}_{RR^{\prime }}[\bullet ] &=&\delta _{R,R^{\prime }}\left\{ 
\frac{-i}{\hbar }[H_{R}^{\prime },\bullet ]-\{D_{R},\bullet
\}_{+}+F_{R}[\bullet ]\right\}  \notag \\
&&+F_{RR^{\prime }}[\bullet ]-\delta _{R,R^{\prime }}\sum\limits_{\substack{ %
R^{\prime \prime }  \\ R^{\prime \prime }\neq R}}\{D_{R^{\prime \prime
}R},\bullet \}_{+},  \label{MComponents}
\end{eqnarray}%
where $H_{R}^{\prime }=H_{R}^{eff}-H_{S},$ is the shift Hamiltonian produced
by the interaction with the reservoir. The initial condition reads $%
\left\vert \tilde{\rho}(0)\right) =\left\vert P\right) \rho _{S}(0),$ where
we have introduced the vector $\left\vert P\right) =(P_{1},P_{2},\ldots
P_{R},\ldots )^{\mathrm{T}}.$ The system state Eq.~(\ref{Suma}) reads $\rho
_{S}(t)=(1\left\vert \tilde{\rho}(t)\right) ,$ where $\left\vert 1\right) $
is the row vector with elements equal to one. Notice that due to the
normalization of the statistical weights it follows $(1\left\vert P\right)
=1.$

From Eq.~(\ref{local}), the system state can be trivially written in the
Laplace domain as 
\begin{subequations}
\begin{eqnarray}
\rho _{S}(u) &=&\left( 1\right\vert \frac{1}{u-(\mathcal{L}_{H}+\mathbb{\hat{%
M}})}\left\vert P\right) \rho _{S}(0), \\
&\equiv &\left( 1\right\vert \mathbb{\hat{G}}(u)\left\vert P\right) \rho
_{S}(0),
\end{eqnarray}
where $u$ is the conjugate variable. Multiplying the right term by the
identity operator written in the form $1/\left( 1\right\vert \mathbb{\hat{G}}%
(u)[u-(\mathcal{L}_{H}+\mathbb{\hat{M}})]\left\vert P\right) ,$ it is
straightforward to arrive to the non-local evolution 
\end{subequations}
\begin{equation}
\frac{d\rho _{S}(t)}{dt}=\mathcal{L}_{H}[\rho _{S}(t)]+\int_{0}^{t}d\tau \,%
\mathbb{L}(t-\tau )[\rho _{S}(\tau )],  \label{memoria}
\end{equation}%
where the superoperator $\mathbb{L}(t)$ is defined by the relation%
\begin{equation}
\left( 1\right\vert \mathbb{\hat{G}}(u)\mathbb{\hat{M}}\left\vert P\right)
[\bullet ]=\left( 1\right\vert \mathbb{\hat{G}}(u)\left\vert P\right) 
\mathbb{L}(u)[\bullet ].  \label{kernel}
\end{equation}
In general, depending on the underlying structure, the evolution Eq.~(\ref%
{memoria}) involves many different memory kernels, each one associated to a
Lindblad contribution.

We notice that a similar master equation was obtained in Refs.~\cite%
{gbma,jpa}. Nevertheless, here the dynamics may strongly departs with
respect to the evolutions that arise from Lindblad equations with a random
rate $[a_{RR^{\prime }}^{\alpha \gamma }=0].$ In fact, the previous
calculation steps are valid only if%
\begin{equation}
\lim {}_{u\rightarrow 0}\left( 1\right\vert u\mathbb{\hat{G}}(u)\left\vert
P\right) =0.  \label{Condicion}
\end{equation}%
By using that $\lim_{t\rightarrow \infty }f(t)=\lim_{u\rightarrow 0}uf(u),$
this condition is equivalent to $\lim {}_{t\rightarrow \infty }\left(
1\right\vert \mathbb{\hat{G}}(t)\left\vert P\right) =0.$ In the general case 
$a_{RR^{\prime }}^{\alpha \gamma }\neq 0,$ Eq.~(\ref{Condicion}) is not
always satisfied. In this situation, the density matrix evolution becomes
non-homogenous and the stationary state may depends on the system initial
condition. In general, this case may arises when the diagonal contributions
are null, i.e., $a_{R}^{\alpha \gamma }=0$ and $a_{RR^{\prime }}^{\alpha
\gamma }\neq 0.$ We remark that these matrix structures values are
completely consistent with the conditions Eq.~(\ref{CondicionesPositividad}%
). On the other hand, in the context of the GBMA, this case arise when the
diagonal sub-bath correlations are null, $\chi _{RR}^{\alpha \beta }(-\tau
)=0,$ which in turn implies that the interaction Hamiltonian Eq.~(\ref{HI})
satisfies $\Pi _{R}H_{I}\Pi _{R^{\prime }}=0$ if $R=R^{\prime }.$

In order to characterize the dynamics when the condition Eq.~(\ref{Condicion}%
) is not satisfied, we introduce the difference 
\begin{subequations}
\begin{eqnarray}
\delta \rho _{S}(u) &\equiv &\rho _{S}(u)-\frac{1}{u}\lim_{u\rightarrow
0}\left( 1\right\vert u\mathbb{\hat{G}}(u)\left\vert P\right) \rho _{S}(0),
\\
&=&\left( 1\right\vert \mathbb{\hat{G}}(u)-\frac{1}{u}\lim_{u\rightarrow 0}u%
\mathbb{\hat{G}}(u)\left\vert P\right) \rho _{S}(0), \\
&\equiv &\left( 1\right\vert \delta \mathbb{\hat{G}}(u)\left\vert P\right)
\rho _{S}(0),
\end{eqnarray}
where now the pseudo-propagator $\delta \mathbb{\hat{G}}(u)$ satisfies $\lim
{}_{u\rightarrow 0}\left( 1\right\vert u\delta \mathbb{\hat{G}}(u)\left\vert
P\right) =0.$ Therefore, $\delta \rho _{S}(t)$ satisfies an evolution like
Eq.~(\ref{memoria}) where the kernel is defined by Eq.~(\ref{kernel}) with $%
\mathbb{\hat{G}}(u)\rightarrow \delta \mathbb{\hat{G}}(u).$ Notice that the
system state, even in the stationary regime, involves the contribution $%
\lim_{u\rightarrow 0}\left( 1\right\vert u\mathbb{\hat{G}}(u)\left\vert
P\right) \rho _{S}(0),$ that in fact depends on the system initial condition.

In the next examples we show the meaning of this property, as well as its
interpretation in the context of the stochastic approach.

\subsection{Dephasing environment}

Here we analyze the case of a qubit system interacting with a dispersive
reservoir \cite{nielsen,budini} whose action can be written in terms of a
dispersive Lindblad rate equation. We assume a complex reservoir with only
two subspaces, $R=a,b,$ whose statistical weights [Eq.~(\ref{pesos})]
satisfy $P_{a}+P_{b}=1.$ Thus, the system state reads 
\end{subequations}
\begin{equation}
\rho _{S}(t)=\tilde{\rho}_{a}(t)+\tilde{\rho}_{b}(t).  \label{sumaAB}
\end{equation}%
A generalization to an arbitrary number of sub-reservoir is straightforward.

The evolution of the auxiliary states are taken as 
\begin{subequations}
\label{auxiliar}
\begin{eqnarray}
\dfrac{d}{dt}\tilde{\rho}_{a}(t) &=&-\gamma _{a}[\tilde{\rho}_{a}(t)-\sigma
_{z}\tilde{\rho}_{a}(t)\sigma _{z}]  \notag \\
&&-\gamma _{ba}\tilde{\rho}_{a}(t)+\gamma _{ab}\sigma _{z}\tilde{\rho}%
_{b}(t)\sigma _{z}, \\
\dfrac{d}{dt}\tilde{\rho}_{b}(t) &=&-\gamma _{b}[\tilde{\rho}_{b}(t)-\sigma
_{z}\tilde{\rho}_{b}(t)\sigma _{z}]  \notag \\
&&-\gamma _{ab}\tilde{\rho}_{b}(t)+\gamma _{ba}\sigma _{z}\tilde{\rho}%
_{a}(t)\sigma _{z},
\end{eqnarray}
where $\sigma _{z}$ is the $z$ Pauli matrix. The completely positive
conditions Eq.~(\ref{CondicionesPositividad}) imply 
\end{subequations}
\begin{subequations}
\label{Rates}
\begin{eqnarray}
\gamma _{a} &\geq &0,\ \ \ \ \ \ \ \ \gamma _{b}\geq 0, \\
\gamma _{ab} &\geq &0,\ \ \ \ \ \ \ \ \gamma _{ba}\geq 0.
\end{eqnarray}

By denoting the matrix elements by $(R=a,b)$%
\end{subequations}
\begin{equation}
\tilde{\rho}_{R}(t)=\left( 
\begin{array}{cc}
\Pi _{R}^{+}(t) & \Phi _{R}^{+}(t) \\ 
\Phi _{R}^{-}(t) & \Pi _{R}^{-}(t)%
\end{array}%
\right) ,  \label{notation}
\end{equation}%
the evolution corresponding to the populations read 
\begin{subequations}
\label{poblacion}
\begin{eqnarray}
\frac{d}{dt}\Pi _{a}^{\pm }(t) &=&-\gamma _{ba}\Pi _{a}^{\pm }(t)+\gamma
_{ab}\Pi _{b}^{\pm }(t), \\
\frac{d}{dt}\Pi _{b}^{\pm }(t) &=&-\gamma _{ab}\Pi _{b}^{\pm }(t)+\gamma
_{ba}\Pi _{a}^{\pm }(t),
\end{eqnarray}%
with $\Pi _{R}^{\pm }(0)=P_{R}\Pi _{S}^{\pm }(0),$ while for the coherences
we obtain 
\end{subequations}
\begin{subequations}
\label{PhiPhi}
\begin{eqnarray}
\frac{d}{dt}\Phi _{a}^{\pm }(t) &=&-(\gamma _{a}+\gamma _{ba})\Phi _{a}^{\pm
}(t)-\gamma _{ab}\Phi _{b}^{\pm }(t), \\
\frac{d}{dt}\Phi _{b}^{\pm }(t) &=&-(\gamma _{b}+\gamma _{ab})\Phi _{b}^{\pm
}(t)-\gamma _{ba}\Phi _{a}^{\pm }(t),
\end{eqnarray}%
with $\Phi _{R}^{\pm }(0)=P_{R}\Phi _{S}^{\pm }(0).$ For expressing the
initial conditions we have trivially extended the notation Eq.~(\ref%
{notation}) to the matrix elements of $\rho _{S}(t).$

We notice that all coherences and populations evolve independently each of
the others. From the evolution of the populations Eq.~(\ref{poblacion}) it
follow 
\end{subequations}
\begin{subequations}
\label{trazas}
\begin{eqnarray}
\frac{d}{dt}\mathrm{Tr}[\tilde{\rho}_{a}(t)] &=&-\gamma _{ba}\mathrm{Tr}[%
\tilde{\rho}_{a}(t)]+\gamma _{ab}\mathrm{Tr}[\tilde{\rho}_{b}(t)],\ \ \ \ \ 
\\
\frac{d}{dt}\mathrm{Tr}[\tilde{\rho}_{b}(t)] &=&-\gamma _{ab}\mathrm{Tr}[%
\tilde{\rho}_{b}(t)]+\gamma _{ba}\mathrm{Tr}[\tilde{\rho}_{a}(t)],\ \ \ \ \ 
\end{eqnarray}%
with $\mathrm{Tr}[\tilde{\rho}_{a}(0)]+\mathrm{Tr}[\tilde{\rho}%
_{b}(0)]=P_{a}+P_{b}=1,$ which implies that the trace of the auxiliary
states perform a classical random walk.

From Eqs.~(\ref{sumaAB}) and (\ref{poblacion}) it becomes evident that the
populations of the system remain unchanged during all the evolution. On the
other hand, the dynamic of the coherences can be obtained straightforwardly
in the Laplace domain. From Eq.~(\ref{PhiPhi}) we get 
\end{subequations}
\begin{equation}
\Phi _{a}^{\pm }(u)=h_{ab}(u)\Phi _{S}^{\pm }(0),\ \ \ \ \ \ \ \Phi
_{b}^{\pm }(u)=h_{ba}(u)\Phi _{S}^{\pm }(0),  \label{PhiAB}
\end{equation}%
where we have introduced the auxiliary function%
\begin{equation}
h_{ab}(u)=\frac{(P_{a}-P_{b})\gamma _{ab}+P_{a}(u+\gamma _{b})}{\gamma
_{ba}(u+\gamma _{a})+\gamma _{ab}(u+\gamma _{b})+(u+\gamma _{a})(u+\gamma
_{b})}.  \label{hab}
\end{equation}%
Therefore, from Eq.~(\ref{sumaAB}) the matrix elements of $\rho _{S}(t)$ read%
\begin{equation}
\Pi _{S}^{\pm }(t)=\Pi _{S}^{\pm }(0),\;\;\;\;\;\;\;\;\Phi _{S}^{\pm
}(t)=h(t)\Phi _{S}^{\pm }(0),  \label{solucion}
\end{equation}%
where $h(t)=h_{ab}(t)+h_{ba}(t),$ gives the coherences decay. From these
solutions, it is straightforward to obtain the corresponding system evolution%
\begin{equation}
\frac{d\rho _{S}(t)}{dt}=\int_{0}^{t}d\tau K(t-\tau )\mathcal{L}[\rho
_{S}(\tau )],
\end{equation}%
with $\mathcal{L[}\bullet ]=(-\bullet +\sigma _{z}\bullet \sigma _{z})$ and $%
K(u)=[1-uh(u)]/h(u).$

In order to check the completely positive condition, we write the solution
map as 
\begin{equation}
\rho _{S}(t)=g_{+}(t)\rho (0)+g_{-}(t)\sigma _{z}\rho (0)\sigma _{z}
\end{equation}%
with $g_{\pm }(t)=[1\pm h(t)]/2.$ This mapping is completely positive at all
times if $g_{\pm }(t)\geq 0$ \cite{petruccione,alicki,nielsen}, and in turn
implies the constraint%
\begin{equation}
|h(t)|\leq 1.  \label{hhhh}
\end{equation}
\begin{figure}[tbh]
\centering
\includegraphics[height=6cm,bb=12 12 301 225]{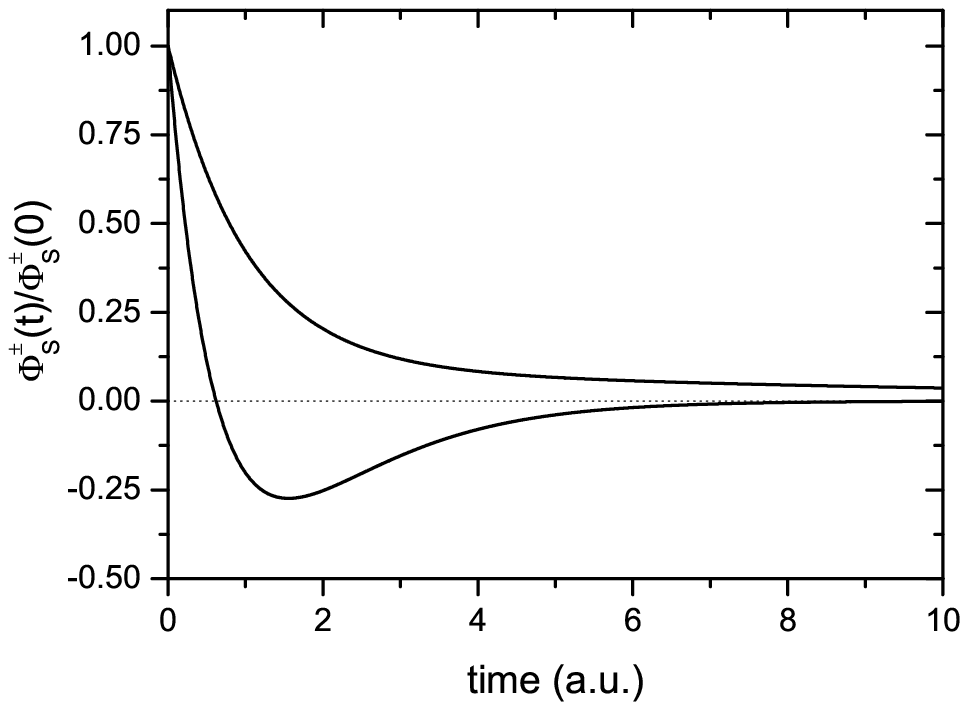}
\caption{Normalized coherences $\Phi _{S}^{\pm }(t)/\Phi _{S}^{\pm
}(0)=h(t), $ Eq.~(\protect\ref{solucion}). In the upper curve the parameters
are $\protect\gamma _{a}=0.1,$ $\protect\gamma _{b}=1,$ and $\protect\gamma %
_{ab}=\protect\gamma _{ba}=0.$ In the lower curve they are $\protect\gamma %
_{a}=0.1,$ $\protect\gamma _{b}=1,$ $\protect\gamma _{ab}=1,$ and $\protect%
\gamma _{ba}=0.1.$ The rates are expressed in arbitrary units (a.u.). In
both curves we take $P_{a}=0.1$ and $P_{b}=0.9.$}
\label{LaUno}
\end{figure}

In the upper curve of Fig.~(\ref{LaUno}) we plot the normalized coherences $%
\Phi _{S}^{\pm }(t)/\Phi _{S}^{\pm }(0)=h(t)$ for the case in which the
non-diagonal rates are null, $\gamma _{ab}=\gamma _{ba}=0.$ Then, the
dynamics reduce to a superposition of exponential decays, each one
participating with weights $P_{a}$ and $P_{b}.$

In the lower curve of Fig.~(\ref{LaUno}) the non-diagonal rates are
non-null, while the rest of the parameters remain the same as in the upper
curve. In contrast to the previous case, here the coherence decay develops
an oscillatory behavior that attain negative values. Clearly, this regime is
unreachable by a superposition of exponential decays.

In both cases, the condition Eq.~(\ref{hhhh}) is satisfied, guaranteeing the
physical validity of the respective solutions.

\subsubsection*{Stochastic representation}

The evolution Eq.~(\ref{auxiliar}) admits a stochastic interpretation like
that proposed previously. The stochastic trajectories can be simulated with
the following algorithms. First, for being consistent with the initial
condition, the\textit{\ system initialization} must be realized as follows

i) Generate a random number $r\in (0,1).$

ii) If $r\leq P_{a}$ $(r>P_{a})$ the dynamic initialize in channel $a$ $(b)$
with $\tilde{\rho}_{a}(0)=\rho _{S}(0)\ $ $[\tilde{\rho}_{b}(0)=\rho
_{S}(0)].$

Trivially, with this procedure the channel $a$ $(b)$ is initialized with
probability $P_{a}$ $(P_{b}).$

By comparing Eqs.~(\ref{auxiliar}) and (\ref{Walk}), the scattering
superoperator results $\mathcal{E}[\bullet ]=\sigma _{z}\bullet \sigma _{z},$
which does not depends on the channel $(a$ and $b).$ It action over an
arbitrary state [Eq.~(\ref{notation})] is $(R=a,b)$ 
\begin{equation}
\mathcal{E}[\tilde{\rho}_{R}(t)]=\sigma _{z}\tilde{\rho}_{R}(t)\sigma
_{z}=\left( 
\begin{array}{cc}
\Pi _{R}^{+}(t) & -\Phi _{R}^{+}(t) \\ 
-\Phi _{R}^{-}(t) & \Pi _{R}^{-}(t)%
\end{array}
\right).  \label{superoperator}
\end{equation}
Therefore, its application implies a change of sign for the coherence
components. On the other hand, the self-dynamics Eq.~(\ref{self}) of each
channel is defined by $\mathcal{\bar{L}}_{a/b}[\bullet ]=\gamma
_{a/b}(-\bullet +\sigma _{z}\bullet \sigma _{z}).$

With the previous information, the \textit{single trajectories} can be
constructed with the following algorithm:

1) Given that the system has arrived at time $t_{i}$ to channel $a,$
generate a random number $r\in (0,1)$ and solve for $(t_{i+1}-ti)$ from the
equation $P_{0}^{(a)}(t_{i+1}-ti)=r,$ where $P_{0}^{(a)}(t)=\exp [-\gamma
_{ba}t].$

2) For times satisfying $t\in (t_{i+1},ti),$ the dynamics in channel $a$ is
defined by its self-propagator, $\tilde{\rho}_{a}(t)=\exp [(t-t_{i})\mathcal{%
\bar{L}}_{a}]\tilde{\rho}_{a}(t_{i}).$

3) At time $t_{i+1}$ the system is transferred from channel $a$ to $b,$
implying the transformation $\tilde{\rho}_{b}(t_{i+1})\rightarrow \mathcal{E}
[\tilde{\rho}_{a}(t_{i+1})]$ and the posterior resetting of channel $a,$
defined by $\tilde{\rho}_{a}(t_{i+1})\rightarrow 0.$

4) Go to 1) with $a\leftrightarrow b$ and $i\rightarrow i+1.$

At this point, it is immediate to realize that the classical rate equations
Eqs.~(\ref{poblacion}) and (\ref{trazas}) arise straightforwardly from the
(transfer) jumps between both channels. The corresponding stationary traces
read 
\begin{equation}
\mathrm{Tr}[\tilde{\rho}_{a}(\infty )]=\frac{\gamma _{ab}}{\gamma
_{ab}+\gamma _{ba}},\ \ \ \ \ \mathrm{Tr}[\tilde{\rho}_{b}(\infty )]=\frac{%
\gamma _{ba}}{\gamma _{ab}+\gamma _{ba}},  \label{poblacionesEstacionarias}
\end{equation}%
which do not depend on the system initial state.

\begin{figure}[tbh]
\centering
\includegraphics[height=6cm,bb= 13 13 300 220]{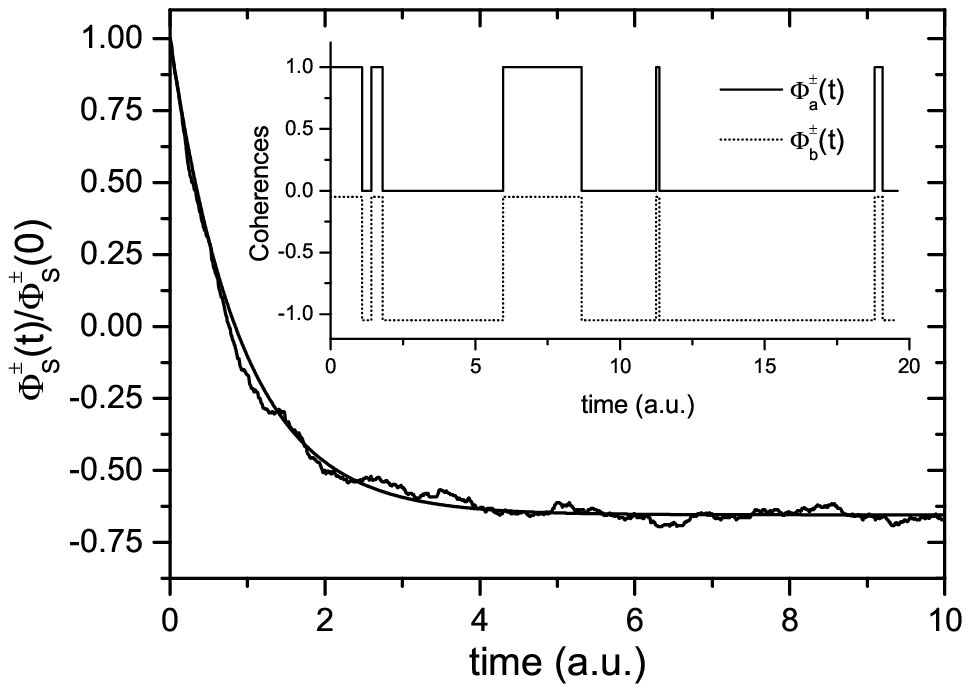}
\caption{Normalized coherences $\Phi _{S}^{\pm }(t)/\Phi _{S}^{\pm
}(0)=h(t), $ Eq.~(\protect\ref{solucion}). The parameters are $\protect%
\gamma _{a}=\protect\gamma _{b}=0,$ $\protect\gamma _{ab}=1,$ $\protect%
\gamma _{ba}=0.1,$ with the statistical weights $P_{a}=0.1$ and $P_{b}=0.9.$
The noisy curve correspond to an average over 500 realizations of the
trajectories defined in the text. The inset show a particular realization
for the coherences $\Phi _{a}^{\pm }(t)$ and $\Phi _{b}^{\pm }(t)$ of the
auxiliary matrixes $\tilde{\protect\rho}_{a}(t)$ and $\tilde{\protect\rho}%
_{b}(t)$ respectively.}
\label{LaDos}
\end{figure}

In contrast with the population evolution, some non-standard dynamical
properties can be found in the coherences evolution when $\gamma _{a}=\gamma
_{b}=0.$ In Fig.~(\ref{LaDos}) we show the normalized coherences $\Phi
_{S}^{\pm }(t)/\Phi _{S}^{\pm }(0)=h(t)$ corresponding to this case. In the
inset, it is shown a typical stochastic realization of the coherences of the
auxiliary matrixes $\tilde{\rho}_{a}(t)$ and $\tilde{\rho}_{b}(t)$ obtained
with the previous algorithm. As expected, in each application of $\mathcal{E}
$ the coherences are transferred between both channels with a change of
sign. We also show an average over 500 realizations. We checked that by
increasing the number of realizations, the average behavior result
indistinguishable with the dynamics Eq.~(\ref{solucion}). 

In strong contrast with the previous figure, in Fig.~(\ref{LaDos}) the
stationary values of the coherences are \textquotedblleft not null and
depend on the initial condition.\textquotedblright\ In fact, their
normalized asymptotic value is $\lim_{t\rightarrow \infty }\Phi _{S}^{\pm
}(t)/\Phi _{S}^{\pm }(0)\simeq -0.654.$ This characteristic is consistent
with the breakdown of condition Eq.~(\ref{Condicion}) and can be understood
in terms of our previous analysis. By taking $\gamma _{a}=\gamma _{b}=0$ in
Eq.~(\ref{PhiAB}) we get%
\begin{equation}
\Phi _{a}^{\pm }(u)=\frac{P_{a}(u+\gamma _{ab})-P_{b}\gamma _{ab}}{%
u[u+\gamma _{ab}+\gamma _{ba}]}\Phi _{S}^{\pm }(0),
\end{equation}%
which implies the asymptotic value 
\begin{subequations}
\label{estacion}
\begin{eqnarray}
\lim_{t\rightarrow \infty }\Phi _{a}^{\pm }(t) &=&(P_{a}-P_{b})\frac{\gamma
_{ab}}{\gamma _{ab}+\gamma _{ba}}\Phi _{S}^{\pm }(0),\ \ \ \  \\
&=&(P_{a}-P_{b})\mathrm{Tr}[\tilde{\rho}_{a}(\infty )]\Phi _{S}^{\pm }(0).
\end{eqnarray}
This last expression can be easily interpreted in terms of the realizations
of the proposed stochastic dynamics. From the inset of Fig.~(\ref{LaDos}),
it is clear that, in spite of a change of sign, the coherence transferred
between both channels does not change along all the evolution. In fact,
notice that due to the election $\gamma _{a}=\gamma _{b}=0,$ the
self-propagators of\ both channels [see previous step 2)] are the identity
operator. Therefore, all realizations that begin in channel $a$ [measured by 
$P_{a}]$ that are found in channel $a$ in the stationary regime (measured by 
$\mathrm{Tr}[\tilde{\rho}_{a}(\infty )]),$ contributes to the stationary
value of the coherence $\Phi _{a}^{\pm }(t)$ with the value $\Phi _{S}^{\pm
}(0).$ This argument explain the contribution proportional to $P_{a}\mathrm{%
Tr}[\tilde{\rho}_{a}(\infty )]\Phi _{S}^{\pm }(0)$ in Eq.~(\ref{estacion}).
On the other hand, a similar contribution is expected from the realizations
that begin in channel $b.$ Nevertheless, due to the action of the
superoperator $\mathcal{E}$ [Eq.~(\ref{superoperator})] they contributes
with the opposite sign.

By adding the contributions of both auxiliary matrixes, from Eq.~(\ref%
{estacion}) the stationary system coherences reads 
\end{subequations}
\begin{equation}
\lim_{t\rightarrow \infty }\Phi _{S}^{\pm }(t)=(P_{a}-P_{b})\left\{ \frac{%
\gamma _{ab}-\gamma _{ba}}{\gamma _{ab}+\gamma _{ba}}\right\} \Phi _{S}^{\pm
}(0)\neq 0,  \label{CoherenciaSistema}
\end{equation}%
This expression fits the stationary value of Fig.~(\ref{LaDos}).

The stochastic realizations corresponding to the system coherence $\Phi
_{S}^{\pm }(t)$ can be trivially obtained from the the realizations of $\Phi
_{a}^{\pm }(t)$ and $\Phi _{b}^{\pm }(t).$ By adding the upper and lower
realizations of the inset of Fig.~(\ref{LaDos}), we get a function that
fluctuates between the values $\pm \Phi _{S}^{\pm }(0).$ By considering the
initial conditions and the superoperator action from these realizations it
is also possible to understand the four contribution terms of Eq.~(\ref%
{CoherenciaSistema}). Finally, we remark that when any of both channels have
a non-trivial self-dynamics, the coherences vanish in the stationary regime,
losing any dependence on the system initial condition $\rho _{S}(0)$ [see
Fig.~(\ref{LaUno})].

\subsection{Depolarizing reservoir}

Another example that admits a stochastic representation is the case of a
depolarizing reservoir \cite{nielsen,budini}, which is defined by the
superoperator 
\begin{equation}
\mathcal{E}[\bullet ]=(\sigma _{x}\bullet \sigma _{x}+\sigma _{y}\bullet
\sigma _{y})/2,  \label{SuperDepol}
\end{equation}%
where $\sigma _{x}$ and $\sigma _{y}$ are the $x$ and $y$ Pauli matrixes
respectively. For simplifying the analysis we assume channels without
self-dynamics. Therefore, the evolution reads 
\begin{subequations}
\begin{eqnarray}
\dfrac{d}{dt}\tilde{\rho}_{a}(t) &=&-\gamma _{ba}\tilde{\rho}_{a}(t)+\gamma
_{ab}\mathcal{E}[\tilde{\rho}_{b}(t)], \\
\dfrac{d}{dt}\tilde{\rho}_{b}(t) &=&-\gamma _{ab}\tilde{\rho}_{b}(t)+\gamma
_{ba}\mathcal{E}[\tilde{\rho}_{a}(t)].
\end{eqnarray}%
The action of the superoperator $\mathcal{E}$ over the states $\tilde{\rho}%
_{R}(t)$ [Eq.~(\ref{notation})] is given by $(R=a,b)$%
\end{subequations}
\begin{equation}
\mathcal{E}[\tilde{\rho}_{R}(t)]=\left( 
\begin{array}{cc}
\Pi _{R}^{-}(t) & 0 \\ 
0 & \Pi _{R}^{+}(t)%
\end{array}%
\right) .  \label{Super}
\end{equation}%
Therefore, its application destroy the coherences components and interchange
the populations of the upper and lower states.

The populations of the auxiliary states evolve as 
\begin{subequations}
\label{poblacionesDepol}
\begin{eqnarray}
\frac{d}{dt}\Pi _{a}^{+}(t) &=&-\gamma _{ba}\Pi _{a}^{+}(t)+\gamma _{ab}\Pi
_{b}^{-}(t), \\
\frac{d}{dt}\Pi _{b}^{-}(t) &=&-\gamma _{ab}\Pi _{b}^{-}(t)+\gamma _{ba}\Pi
_{a}^{+}(t),
\end{eqnarray}%
subject to the initials conditions $\Pi _{a}^{+}(0)=P_{a}\Pi _{S}^{+}(0)$
and $\Pi _{b}^{-}(0)=P_{b}\Pi _{S}^{-}(0).$ The evolution of $\Pi
_{b}^{+}(t) $ and $\Pi _{a}^{-}(t)$ follows after changing $a\leftrightarrow
b.$ Notice that this splitting of the population couplings follows from the
superoperator action defined by Eq.~(\ref{Super}). On the other hand, the
coherences evolution read 
\end{subequations}
\begin{equation}
\frac{d}{dt}\Phi _{a}^{\pm }(t)=-\gamma _{ba}\Phi _{a}^{\pm }(t),\ \ \ \ \ 
\frac{d}{dt}\Phi _{b}^{\pm }(t)=-\gamma _{ab}\Phi _{b}^{\pm }(t).
\end{equation}%
Therefore, in this case the stationary coherences are null. This fact also
follows trivially from Eq. (\ref{Super}). In contrast, the stationary
populations reads 
\begin{subequations}
\begin{eqnarray}
\Pi _{a}^{+}(\infty ) &=&[\Pi _{S}^{+}(0)P_{a}+\Pi _{S}^{-}(0)P_{b}]\frac{%
\gamma _{ab}}{\gamma _{ab}+\gamma _{ba}}, \\
\Pi _{b}^{-}(\infty ) &=&[\Pi _{S}^{+}(0)P_{a}+\Pi _{S}^{-}(0)P_{b}]\frac{%
\gamma _{ba}}{\gamma _{ab}+\gamma _{ba}},
\end{eqnarray}%
where $\Pi _{b}^{+}(\infty )$ and $\Pi _{a}^{-}(\infty )$ follows after
changing $a\leftrightarrow b.$ This result has an immediate interpretation
in the context of the stochastic approach. In fact, the last fractional
factors correspond to the \textquotedblleft natural\textquotedblright\
stationary solutions of Eq.~(\ref{poblacionesDepol}). This solution is
corrected by the terms in brackets, which in fact take in account the system
initialization [notice that $\Pi _{a}^{+}(0)+\Pi _{b}^{-}(0)\neq 1]$ and the
transformations induced by the superoperator $\mathcal{E}$ Eq.~(\ref{Super}%
). Finally, the system stationary populations $\Pi _{S}^{\pm }(\infty )=\Pi
_{a}^{\pm }(\infty )+\Pi _{b}^{\pm }(\infty )$ reads 
\end{subequations}
\begin{equation}
\Pi _{S}^{\pm }(\infty )=\Pi _{S}^{\pm }(0)\frac{P_{a}\gamma
_{ab}+P_{b}\gamma _{ba}}{\gamma _{ab}+\gamma _{ba}}+\Pi _{S}^{\mp }(0)\frac{%
P_{a}\gamma _{ba}+P_{b}\gamma _{ab}}{\gamma _{ab}+\gamma _{ba}}.
\end{equation}%
As in the previous case, the dependence of the stationary state in the
initial conditions is lost when the channels have a proper dissipative
self-dynamics.

\section{Summary and Conclusions}

We have presented a new class of dynamical master equations that provide an
alternative framework for the characterization of non-Markovian open quantum
system dynamics. In this approach, the system state is written in terms of a
set of auxiliary matrixes whose evolutions involve Lindblad contributions
with coupling between all of them, resembling the structure of a classical
rate equation.

We have derived the previous structure from different approaches. In the
context of the GBMA, a complex structured reservoir is approximated in terms
of a direct sum of Markovian sub-reservoirs. Then, the Lindblad rate
structure arises by considering arbitrary interaction\ Hamiltonians that
couple the different subspaces associated to each sub-reservoir. The matrix
structures that define the system evolution are expressed in terms of the
projected bath correlations.

On the other hand, we have derived the same structure from composite
environments, where the entanglement between the system and a Markovian
environment is modulated by extra unobserved degrees of freedom. The
Lindblad rate structure arises straightforwardly when the tripartite
interaction Hamiltonian that involve the three parts does not couple the
coherences and populations of the extra degrees of freedom. This scheme also
allows to find the conditions under which an arbitrary Lindblad rate
equation provides a completely positive evolution.

Due to the apparent similarity of the evolution with a classical rate
equation, we have also formulated a quantum stochastic dynamics that in
average is described by a Lindblad rate equation. The stochastic dynamic
consists in a set of transmission channels, each one endowed with a
different self-system evolution, and where the transitions between them are
attended by the application of a completely positive superoperator. This
formalism allows to understand some amazing properties of the non-Markovian
dynamics, such as the dependence of the stationary state in the initial
conditions. This phenomenon arise from the interplay between the initial
channel occupations and the structure of the stochastic dynamics. We
exemplified our results by analyzing the dynamical action of non-trivial
complex dephasing and depolarizing reservoirs over a single qubit system.

In conclusion, we have presented a close formalism that defines an extra
class of non-Markovian quantum processes that may be of help for
understanding different physical situations where the presence of non-local
effects is relevant \cite%
{barkaiChem,schlegel,brokmann,grigolini,makhlinReport,falci,rapid,john,quang}%
.

\section*{Acknowledgments}

This work was partially supported by Secretar\'{\i}a de Estado de
Universidades e Investigaci\'{o}n, MCEyC, Spain. The author also thanks
financial support from CONICET, Argentine.

\appendix


\begin{thebibliography}{99}
\bibitem{cohen} C. Cohen-Tannoudji, J. Dupont-Roc, and G. Grynberg, \textit{%
Atom-photon interactions} (Wiley, New York, 1992).

\bibitem{petruccione} H.P. Breuer and F. Petruccione, \textit{The Theory of
Open quantum Systems} (Oxford University Press, Oxford, 2002).

\bibitem{alicki} R. Alicki and K. Lendi, \textit{Quantum Dynamical
Semigroups and Applications}, Lecture Notes in Physics \textbf{286}
(Springer, Berlin, 1987).

\bibitem{nielsen} M.A. Nielsen and I.L. Chuang, \textit{Quantum Computation
and Quantum Information}, (Cambridge University Press, Cambridge, England,
2000).

\bibitem{weiss} U. Weiss, \textit{Quantum Dissipative Systems}, (World
Scientific, 1999).

\bibitem{imamoglu} I. Imamoglu, Phys. Rev. A \textbf{50}, 3650 (1994).

\bibitem{garraway} B.M. Garraway, Phys. Rev. A \textbf{55}, 2290 (1997); 
\textbf{55}, 4636 (1997).

\bibitem{tannor} C. Meier and D.J. Tannor, J. Chem. Phys. \textbf{111}, 3365
(1999).

\bibitem{ulrich} U. Kleinekath\"{o}fer, J. Chem. Phys. \textbf{121}, 2505
(2004).

\bibitem{esposito} M. Esposito and P. Gaspard, Phys. Rev. E \textbf{68},
066112 (2003); \textbf{68}, 066113 (2003).

\bibitem{haake} F. Haake, in \textit{Statistical Treatment of Open Systems
by Generalized Master Equations}, (Springer, 1973).

\bibitem{breuer} H.P. Breuer, J. Gemmer, and M. Michel, Phys. Rev. E \textbf{%
73}, 016139 (2006).

\bibitem{barnett} S.M. Barnett and S. Stenholm, Phys. Rev. A \textbf{64},
033808 (2001).

\bibitem{wilkie} J. Wilkie, Phys. Rev. E \textbf{62}, 8808 (2000).

\bibitem{budini} A.A. Budini, Phys. Rev. A \textbf{69}, 042107 (2004).

\bibitem{cresser} S. Daffer, K. Wodkiewicz, J.D. Cresser, and J.K. McIver,
Phys. Rev. A \textbf{70}, 010304(R) (2004).

\bibitem{lidar} A. Shabani and D.A. Lidar, Phys. Rev. A \textbf{71},
020101(R) (2005).

\bibitem{sabrina} S. Maniscalco, Phys. Rev. A \textbf{72}, 024103 (2005).

\bibitem{maniscalco} S. Maniscalco and F. Petruccione, Phys. Rev. A \textbf{%
73}, 012111 (2006).

\bibitem{wilkieChem} J. Wilkie, J. Chem. Phys. \textbf{114}, 7736 (2001); 
\textit{ibid} \textbf{115}, 10335 (2001).

\bibitem{jpa} A.A. Budini and H. Schomerus, J. Phys. A \textbf{38}, 9251,
(2005).

\bibitem{gbma} A.A. Budini, Phys. Rev. E \textbf{72}, 056106 (2005); e-print
quant-ph/0601140.

\bibitem{salo} J. Salo, S.M. Barnett, and S. Stenholm, Op. Comm. \textbf{259}%
, 772 (2006).

\bibitem{barkaiChem} E. Barkai, Y. Jung, and R. Silbey, Annu. Rev. Phys.
Chem. \textbf{55}, 457 (2004).

\bibitem{schlegel} G. Schlegel, J. Bohnenberger, I. Potapova, and A. Mews,
Phys. Rev. Lett. \textbf{88}, 137401 (2002).

\bibitem{brokmann} X. Brokmann, J.P. Hermier, G. Messin, P. Desbiolles, J.P.
Bouchaud, and M. Dahan, Phys. Rev. Lett. \textbf{90}, 120601 (2003).

\bibitem{grigolini} G. Aquino, L. Palatella, and P. Grigolini, Phys. Rev.
Lett. \textbf{93}, 050601 (2004).

\bibitem{rapid} A.A. Budini, Phys. Rev. A \textbf{73}, 061802(R) (2006).

\bibitem{makhlinReport} Y. Makhlin, G. Sch\"{o}n, and A. Shnirman, Rev. Mod.
Phys. \textbf{73}, 357 (2001).

\bibitem{falci} G. Falci, A. D'Arrigo, A. Mastellone, and E. Paladino, Phys.
Rev. Lett. \textbf{94}, 167002 (2005).

\bibitem{john} S. John and T. Quang, Phys. Rev. Lett. \textbf{74}, 3419
(1994).

\bibitem{quang} T. Quang, M. Woldeyohannes, S. John, and G.S. Agarwal, Phys.
Rev. Lett. \textbf{79}, 5238 (1997).

\bibitem{kampen} N. G. van Kampen, in \textit{Stochastic Processes in
Physics and Chemistry}, 2nd ed. (North-Holland, Amsterdam, 1992).

\bibitem{bariloche} A.A. Budini, A.K. Chattah, and M.O. C\'{a}ceres, J. Phys
A: Math. Gen. \textbf{32}, 631 (1999).
\end{thebibliography}
\end{document}